# Interface-engineered oxidation-resistant wafer-level Tantalum–Tantalum thermocompression bonding for 3D integration of superconducting interconnects

Harsh Mishra, Ullas Pandey, Sathish Bonam, Adirae Dinesh, Sai Rama Krishna Malladi, and Shiv Govind Singh*[1]

***Abstract*— Wafer-level 3D integration of superconducting interconnects requires low-thermal budget bonding processes compatible with superconducting materials. Rapid native oxide formation on tantalum (Ta) surfaces limits low-temperature, low-pressure direct Ta–Ta thermocompression bonding. In this study, we develop an oxidation-resistant bonding process using an ultrathin Au passivation layer to suppress oxide formation during bonding. The engineered interface enables blanket Ta–Ta wafer bonding at 300 °C under 4.93 bar, significantly reducing the bonding thermal budget and generation of α-Ta across the interface, which potentially improves coherence time as reported in literature. Structural and interfacial analyses confirm oxide suppression and continuous metallic bonding, having a bond strength of 169 MPa. This work demonstrates a low-temperature, low-pressure Ta–Ta thermocompression bonding strategy for scalable 3D superconducting interconnect integration.**

***Index Terms*—Oxidation resistive, Thermo-compression Bonding, Tantalum (Ta), 3D integration, Quantum Computing.**

## I. INTRODUCTION

Superconducting quantum computing has progressed from few-qubit demonstrations to complex 3D integrated multi-die processors [1–3]. Such structures use superconducting through silicon vias (TSVs), which minimize resistive losses and thermal dissipation at cryogenic temperatures compared to conventional copper interconnects. Quantum processors function at temperatures below the superconducting critical temperature ($T_c$), where electrical resistance becomes zero. While copper integration has been extensively explored, the large-scale integration of superconducting materials remains a significant challenge [4–8]. Superconducting materials such as Al, Nb, Ta, and TiN have been widely studied [9–17], with tantalum gaining recent attention for its exceptional stability and maintaining high qubit coherence times. In 2022, Ta-based qubits fabricated using dry etching achieved coherence times of 0.5 milliseconds, outperforming Nb and Al-based devices. This enhancement is due to the stability of tantalum's single-phase oxide ($Ta_2O_5$), in contrast to the presence of multiple niobium oxides and the instability of aluminum oxide, both of which lead to increased two-level system (TLS) losses [18-19]. Similarly, in our earlier study, we examined the distribution of Cooper pairs in superconducting interconnects made of Nb, Ta, and annealed Ta under magnetic fields. Ta-based interconnects exhibited a uniform Cooper-pair density compared with Nb-based interconnects [20-21].

Similarly, in multichip 3D integration of superconducting circuits where two chips are joined via thermocompression bonding in vacuum, based on this, we demonstrated the successful wafer-level integration of superconducting interconnects using tantalum (Ta) as a superconducting material [22]. However, Ta-Ta thermocompression bonding requires high temperature due to surface oxidation. To overcome this, we propose Au passivation of Ta films. Gold (Au) has a smaller atomic volume (10.21 cm³/mol) compared to Ta (10.85 cm³/mol) [23], which supports interdiffusion during bonding, creating a void-free interface. Selection of a thin Au-capping layer is based on the theoretical study discussed in section II, which prevents Ta oxidation, reduces surface roughness, and significantly lowers bonding temperature (from 500 °C to 300 °C) while improving joint reliability. Prior studies have also demonstrated Nb interconnects fabricated using Au–Au thin-film thermocompression bonding, in which a multilayer stack consisting of 100 nm Nb, a 5 nm Ir interlayer for oxide suppression, and a 200 nm Au top layer was employed. Following thin-film optimization, thermocompression bonding was performed at a temperature of 140 °C, with a bonding time of 1.5 seconds under a pressure of 35 MPa. The relatively large Au thickness enables low-temperature

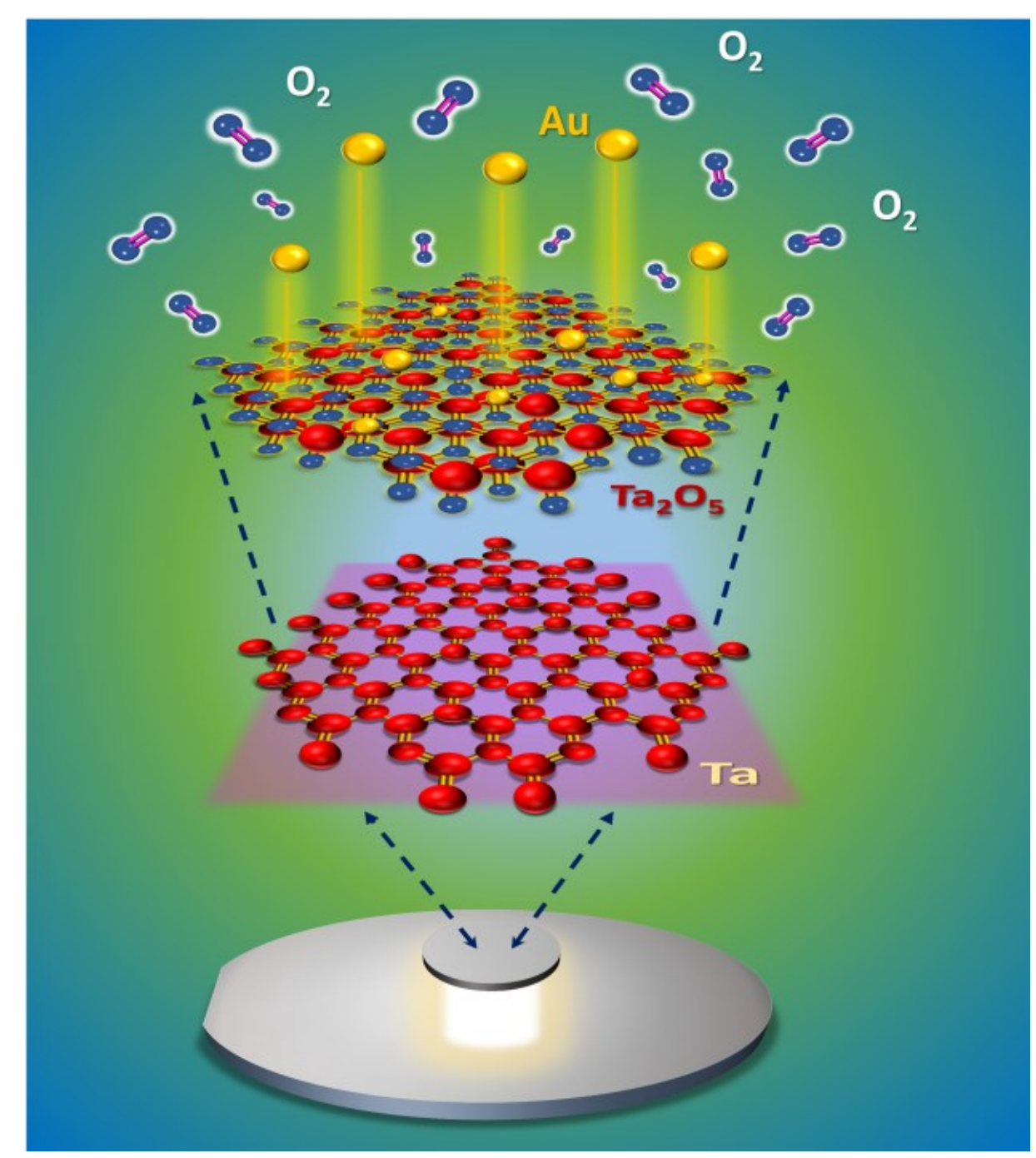

Figure 1. Schematic illustration of Au passivation on Ta, depicting the mechanism of suppression of Tantalum oxide.

Harsh Mishra, Ullas Pandey, and Shiv Govind Singh are with the Department of Electrical Engineering at the Indian Institute of Technology Hyderabad, Sangareddy 502284, India (e-mail: sgsingh@ee.iith.ac.in).
Sathish Bonam is with Tyndall National Institute, Lee Maltings Complex Dyke Parade, Cork, Ireland.
Adirae Dinesh and Sai Rama Krishna Malladi are with the Department of Materials Science and Metallurgical Engineering, Indian Institute of Technology Hyderabad, Sangareddy 502284, India.

bonding, and thus the process is classified as Au–Au thermocompression bonding rather than direct Nb–Nb bonding. Electrical characterization showed that Nb interconnects incorporating Au joints retained a superconducting transition temperature of approximately 9 K, comparable to that of the underlying Ir/Nb stack, indicating that the sandwiched Au layer does not degrade the superconducting critical temperature after bonding [24]. Consistent with these results, recent studies have shown that Au used as a passivation layer on Ta remains superconducting due to proximity coupling with the underlying Ta. Microwave loss measurements showed performance matching that of state-of-the-art Ta circuits at millikelvin temperatures [25]. Furthermore, Au capping of Nb films has been reported to systematically improve average qubit relaxation times, highlighting that noble metal encapsulation can preserve and, in some cases, enhance quantum device performance [26].

Optimizing Au thickness is critical because a too-thick layer hinders diffusion, while a too-thin layer fails to passivate effectively. This strategy provides a chemically inert, low-temperature, and robust bonding pathway for next-generation superconducting quantum processors. A detailed explanation of the selection of passivation material and the proposed bonding mechanism, along with its associated characterizations, can be found in sections II, III, and IV. Figure 1 provides a schematic illustration of the passivation, depicting the oxide suppression.

## II. SELECTION OF CAPPING LAYER

The criterion for selecting suitable capping materials is the existence of intrinsic bulk affinity for oxygen in superconducting metals, which should be minimized to prevent oxidation. This property can be studied through the interstitial formation energy [27], expressed as

$$E_{Inter}^{X} = E_{bulk+X@int} - E_{bulk} - \frac{1}{2}E_{x_2}, \qquad X = O, N, H$$

A positive value of $E_{Inter}^{X}$ shows that the impurity is thermodynamically unfavorable and will tend to decline from the material. Based on this, a subset of noble metals, such as Ag, Au, Pd, Rh, Ru, Ir, and W, showed positive $E_{Inter}^{X}$ values of 0.05, 0.64, 0.28, 1.57, 0.77, 1.77, and 0.38 eV, respectively [27], indicating their resistance to oxygen incorporation. In addition to oxidation resistance, the effect of the capping layer on the superconducting properties must be considered, specifically the proximity effect in superconductor–normal metal (S–N) bilayers, where S represents Ta or Nb and N denotes the passivating layer. Within the framework based on the strong-coupling Eliashberg formalism, and considering the assumption that both layers are thinner than the superconducting coherence length, the superconducting order parameter can be regarded as spatially uniform within each layer. The coupling between the two layers occurs exclusively through the interface, and the effective coupling constants characterize the interfacial proximity strength $\Gamma_S$ and $\Gamma_N$. The ratio between these coupling parameters is approximately given by

$$\frac{\Gamma_S}{\Gamma_N} \approx \frac{d_N N_N(0)}{d_S N_S(0)} ,$$

where $N_{N,S}(0)$ Respective Fermi level densities represent the electronic densities of states at the Fermi level for the respective materials. Thus, the thickness ratio $\frac{d_N}{d_S}$ directly governs the proximity strength and consequently affects the superconducting transition temperature ($T_c$). An increase in $\frac{d_N}{d_S}$ leads to a pronounced suppression of $T_c$. Considering material with low $\frac{d_N}{d_S}$ Noble metals such as Au, Pd, and Pt demonstrate comparable performance as effective passivation layers [27].

## III. OVERVIEW OF BONDING MECHANISM

The experiment employed 2-inch, P-type silicon prime wafers with <100> orientation. These wafers were first cleaned using the standard RCA and Piranha procedure and subjected to a dehydration bake and nitrogen drying. Thin films were deposited sequentially on the prepared silicon wafers, starting with a 175 nm tantalum layer, followed by a thin gold layer (serving as a passivation layer for tantalum) with thicknesses ranging from 1 nm to 7 nm to identify the optimal passivation thickness. All depositions were carried out at room temperature using a DC sputtering system (AJA International Inc., USA) with a base pressure of below $2\times10^{-9}$ Torr, Ar pressure of 3 milliTorr, and power (125W and 75W) for Ta and Au, respectively. The two metal-coated wafers were placed in the bonding chamber with their coated surfaces facing each other. Thermocompression bonding of the metal-coated wafers was performed using an AML bonder (Applied Microengineering Limited, AML-AWB-04), which provides alignment capabilities for wafers up to 4 inches in diameter. The bonding process was conducted at 300°C under a pressure of 4.93 bar for 120 minutes, as shown in Fig. 2.

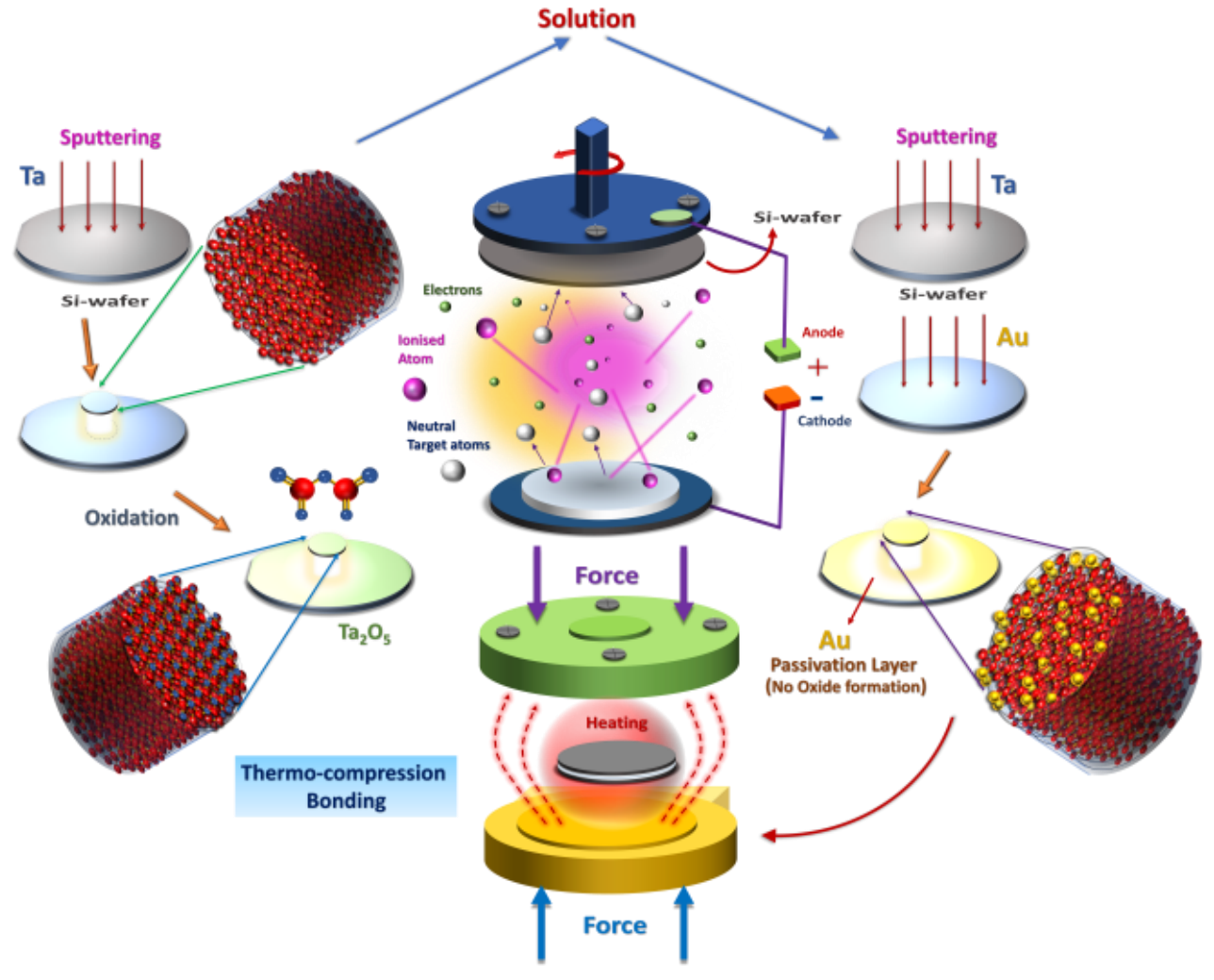

Figure 2. Gold (Au) passivated Tantalum-Tantalum thermocompression bonding process.

## IV. RESULTS AND DISCUSSION

### A. *Surface analysis after Au passivation before wafer bonding*

The optimal thickness of the passivation layer depends on the resulting surface roughness, as a smoother surface requires lower bonding pressure for effective bonding between two wafers. To assess the surface roughness, we conducted an in-depth analysis of the surface roughness of Au-passivated samples as a function of varying passivation layer thickness (1 nm to 7 nm) using atomic force microscopy (Bruker Icon ScanAsyst system). Imaging was carried out in tapping mode over a 64 µm² scan area in each analysis. This method helps us to capture the detailed surface profile and understand how varying thicknesses of Au passivation affect surface roughness at the nanoscale. To establish a baseline for comparison, we evaluated the surface roughness of a control Ta film without any Au passivation. This unpassivated Ta surface exhibited a RMS (Rq) roughness of 0.183 nm, as shown in Fig. 3(a), giving a reference point against which the effects of Au passivation could be measured. The data indicate that applying the Au passivation layer initially leads to a decrease in RMS roughness of 0.177 nm, as shown in Fig. 3(b), thereby smoothing the Ta surface and improving bonding quality at lower pressure. A unique trend observed at higher passivation thicknesses, as the Au layer achieves 5 nm in thickness, complementary RMS roughness begins to increase to 0.202 nm and 0.257 nm for 7 nm, respectively, as shown in Fig. 3(c) and Fig. 3(d), which is likely due to changes in the layer's structural characteristics that may affect uniformity. Among the thickness variations analysis, the 3-nm Au passivation layer shows an RMS roughness of 0.177 nm, suggesting it is a promising thickness for achieving minimal roughness.

Chemical analysis using X-ray photoelectron spectroscopy (AXIS Ultra) was performed for elemental analysis of the Au-passivated samples, focusing on those with Au passivation layer thicknesses of 0, 1, 3, 5, and 7 nm. Figure 4(a) displays the Ta 4f and O 1s photoelectron spectra at the surface of a typical $Ta_2O_5$ film, with the solid line indicating the recorded data. Two notable peaks at 28.9 ± 0.05 eV and 26.9 ± 0.05 eV correspond to the $Ta_2O_5$ markers for Ta $4f_{5/2}$ and Ta $4f_{7/2}$, respectively [28]. For a passivation layer of 1 nm, there is a reduction in peak intensity between 28.9 ± 0.05 eV and 26.9 ± 0.05 eV. Still, the surface isn't fully passivated (as shown in Fig. 4(a)), suggesting that a 1 nm thickness may not provide desirable protection against oxidation. However, we observe a Tantalum metal peak at 3 nm, but no oxide peaks, as shown in the zoomed image in Fig. 4(a), suggesting more effective surface passivation at this thickness. Thus, a minimum of 3 nm of Au passivation appears necessary to minimize oxidation of the Ta surface. For both 3 nm and thicker Au layers, a reduction in the intensity of the underlying Ta peaks shows the increased surface coverage by the Au layer. As shown in Figure 4(b), the intensity of the Au $4f_{7/2}$ (~84 eV) and Au $4f_{5/2}$ (~87.8 eV) peaks rises as the Au thickness increases from 1 to 7 nm [29]. The absence of chemical shifts or additional features confirms that the Au layer remains chemically stable without interfacial reactions with Ta.

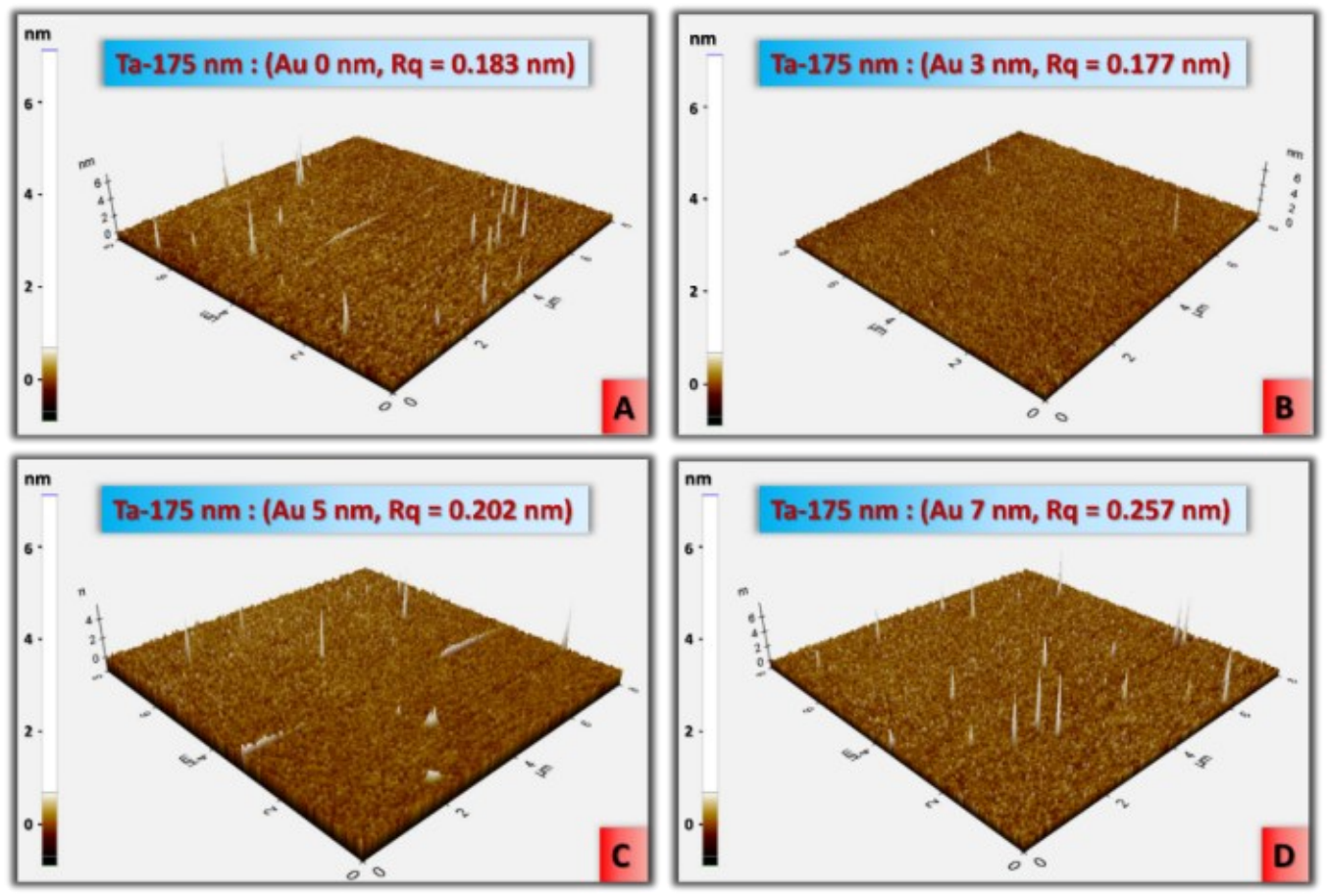

Figure 3. (A) 3D imaging of pure Ta extracted from AFM with RMS roughness of 0.183 nm, (B) 3 nm Au-passivated Ta with RMS roughness of 0.177 nm, (C) 5 nm Au-passivated Ta with RMS roughness of 0.202 nm, (D) 7 nm Au-passivated Ta with RMS roughness of 0.257 nm.

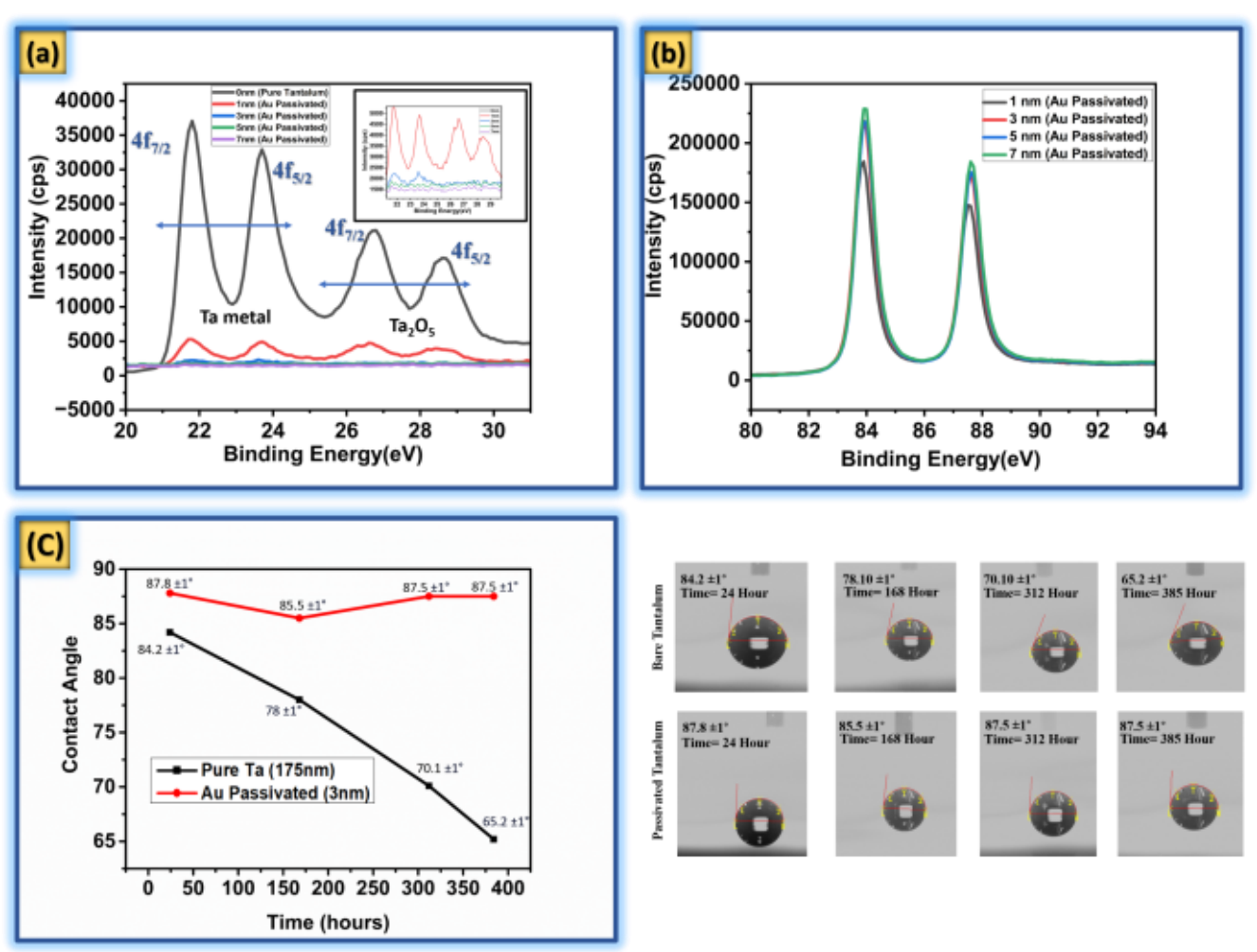

Figure 4. XPS analysis of Au-passivated samples showing elemental scans for (A) tantalum (Ta) and (B) gold (Au). (C) A comparison of contact angle measurements on passivated and bare Ta surfaces, including time-resolved camera images captured during the measurement for both bare and passivated tantalum films.

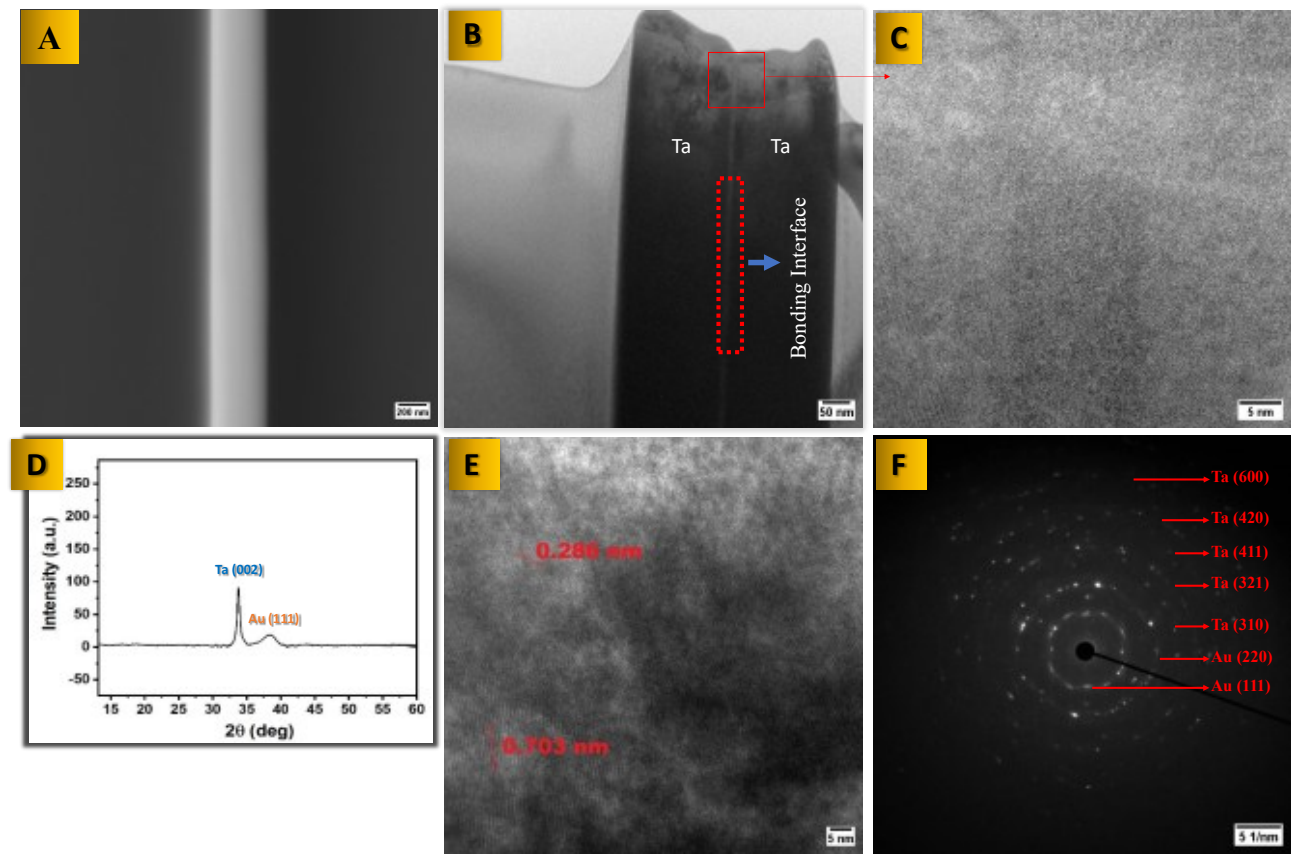

Figure 5. Cross-sectional analysis of the bonded interface: (A) FE-SEM image, (B) TEM lamella showing the full bonding interface, (C) zoomed view of Au-passivated Ta-Ta interface, (D) XRD of Ta thin film with 3 nm Au passivation, (E) HRTEM showing significant grain growth with fringe widths of 0.286 nm and 0.703 nm, (F) SAED pattern displaying Debye–Scherrer rings from Au (111, 220) and Ta (310, 321, 411, 420, 600).

A thicker Au passivation layer generally offers better protection, as confirmed by elemental analysis. However, increasing the thickness too much can limit the diffusion of Au atoms across the bonding interface, which can be detrimental to the bonding quality. A key requirement for any passivation technique is its ability to prevent surface oxidation over long periods. Long-term surface stability is essential, especially in semiconductor manufacturing, where wafers are handled across various locations and technologies. Reliable passivation allows wafers to be prepared independently without concern for surface degradation, enabling smooth wafer-on-wafer (WoW) bonding for 3D integration. This stability becomes even more important in heterogeneous integration, where layers from different technological platforms must align perfectly to build advanced, multi-layered semiconductor architectures. To verify the effectiveness of our Au passivation layer, we performed contact angle measurements using the VCA Optima system (AST Products, Inc.). This method accurately tracks surface oxidation by monitoring changes in surface wettability over time. Contact angle analysis is particularly useful because surface oxidation typically alters the surface chemistry in a way that affects bonding reliability. For example, pure Ta surfaces are naturally hydrophobic, but they become hydrophilic upon oxidation. As shown in Fig. 4(c), the unpassivated (control) Ta sample exhibited a clear drop in contact angle from 84.2° to 65.2°, signaling a transition from hydrophobic to hydrophilic, consistent with earlier studies [30–31]. In contrast, the Au-passivated Ta surface maintained a stable contact angle throughout the measurement period. This stability confirms that the ultrathin Au passivation layer effectively suppresses oxidation and preserves the surface condition over time. This stable contact angle on the Au-passivated sample provides compelling evidence of the passivation layer's reliability in preserving the Ta surface's original properties and enhancing long-term stability under operational conditions.

### B. *Interface and Reliability Analysis After Wafer Bonding (Wafer-Level Ta-Ta Bonding):*

Post-bonding process, interface analysis was carried out using bond strength measurements, cross-sectional imaging by FESEM and TEM, and an evaluation of bond strength reliability test after seven days (pause between deposition and bonding). The primary assessment of bonding interface quality was performed using a simple razor test, in which reliable bonding was confirmed by the inability to insert a razor blade into the bonded interface. One of the samples was also examined under high-resolution scanning electron microscopy (X-FESEM). The X-FESEM image analysis in Fig. 5(A) revealed seamless integration between the two Au-passivated Ta-coated layers, with no visible interface or voids, suggesting a continuous Ta layer across the Ta-Ta bonded interface. This sample does not show a clear boundary between the bonded Ta thin films, suggesting possible recrystallization at the bonded surface interface. To study the grain growth across the bonded interface, the sample was meticulously examined using a state-of-the-art Transmission Electron Microscope (TEM) JEOL JEM-F200, while samples were prepared using FIB on JEOL JIB-4700F from the bonded sample. Cross-sectional TEM (X-TEM) images of the bonded wafers are presented in Fig. 5(B). A high-resolution image of the interface region, shown in Fig. 5(C), reveals that the interface layer within the tantalum bulk region is free of voids. Initially, to confirm the orientation of the Au-passivated tantalum thin film, we carried out XRD analysis, which revealed the presence of Ta (002) and Au (111) planes, as shown in Fig. 5(D) [32]. The HRTEM image in Fig. 5(E) shows significant grain growth with no distinguishable interface, which confirms a seamless, void-free bond essential for achieving high-quality bonding. The observed d-spacings of 0.286 nm and 0.703 nm indicate the presence of multiple crystallographic orientations which is confirmed by rings observed in the selected area electron diffraction (SAED) pattern as shown in the inset in Fig. 5(F), identify the Au (111) and Au (220) planes along with α-Ta planes (310), (321), (411), (420), and (600) and are consistent with prior studies involving high-temperature Ta thin film deposition [33-34]. To further study the presence of gold at the bonding interface after the bonding process, TEM-EDS analysis was conducted on the selected region of the TEM micrograph, as shown in Fig. 6(A). The corresponding elemental maps for Ta, Au, and Si in the Ta-Ta bonded sample are shown in Fig. 6(B), which confirm the presence of Au across the examined area. The line-scan EDS profile in Fig. 6(C) and post-bonding EDX spectra of the bonded structure, shown in Fig. 6(D), indicate a consistent presence of Au throughout the sample, with noticeable diffusion into the Ta layers and the highest intensity observed at the bonding interface. This suggests that the ultrathin gold film spread and integrated throughout the bonded Ta-Ta structure during the bonding process. The applied thermal and pressure conditions likely facilitated this diffusion and redistribution, contributing to the formation of a robust bonding interface.

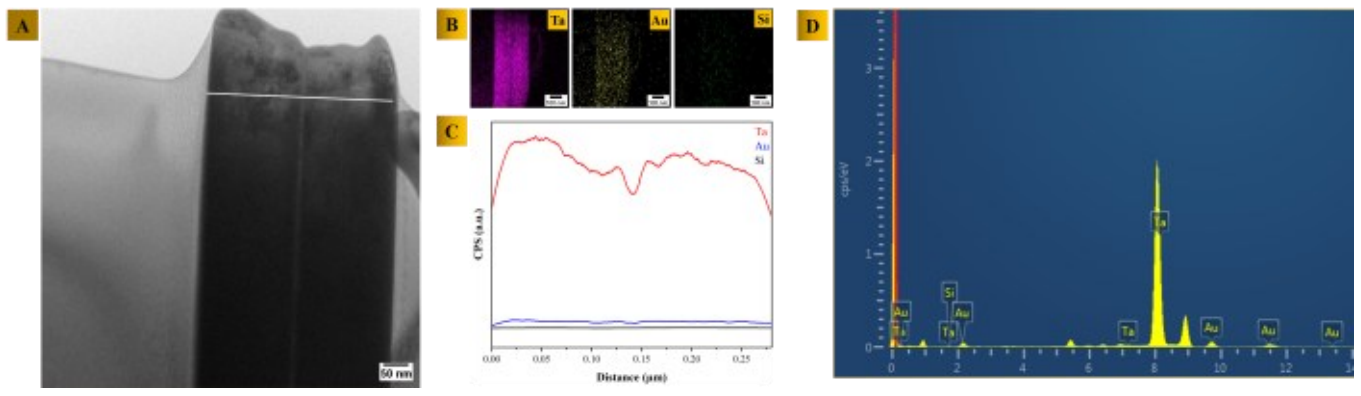

Figure 6. Cross-sectional TEM-EDS analysis of Au-passivated Ta-Ta bonded sample: (A) TEM image showing the analyzed area, (B) EDS elemental maps of Ta, Au, and Si at the interface, (C) line scan showing Au distribution, (D) post-bonding EDX spectra confirming Au presence.

Bond strength is an important measure of bonding quality between wafers. To assess the bond strength of the 2-inch bonded wafer, the sample was diced into small (1 cm × 1 cm) pieces with a diamond cutter. The bonded wafer maintained its integrity after dicing, as shown in Fig. 7(B), indicating strong bonding quality. Shear tests were then performed on these diced pieces using the Instron Micro tester system (model no. 5948 R 504). Each diced specimen was fixed in the specimen holder with specialized screws, and a maximum load of 1000 N was applied to assess the bonded sample's strength, as depicted in Fig. 7(A). The bond strength data for one diced piece is presented in Fig. 7(C), showing a bond strength of 474 N (~ 169 MPa), consistent with previous studies and demonstrating reliable bonding across multiple samples [4-6]. A comparison of bond strength after a seven-day pause is shown in Fig. 7(D). The results indicate that the bond strength of 451 N (~161 MPa) for the Au-passivated Ta wafer aligns with the strength of the Au-passivated bonded wafer with no pause, as seen in Fig. 7(C). Further analysis was conducted on a seven-day pause case without passivation Ta sample, which showed a significant reduction in bond strength to 55.47 N (~19.88 MPa), as shown in Fig. 7(E), compared to the paused, Au-passivated Ta-bonded wafers. This reduction is likely due to the oxide formation, which impacts bond strength.

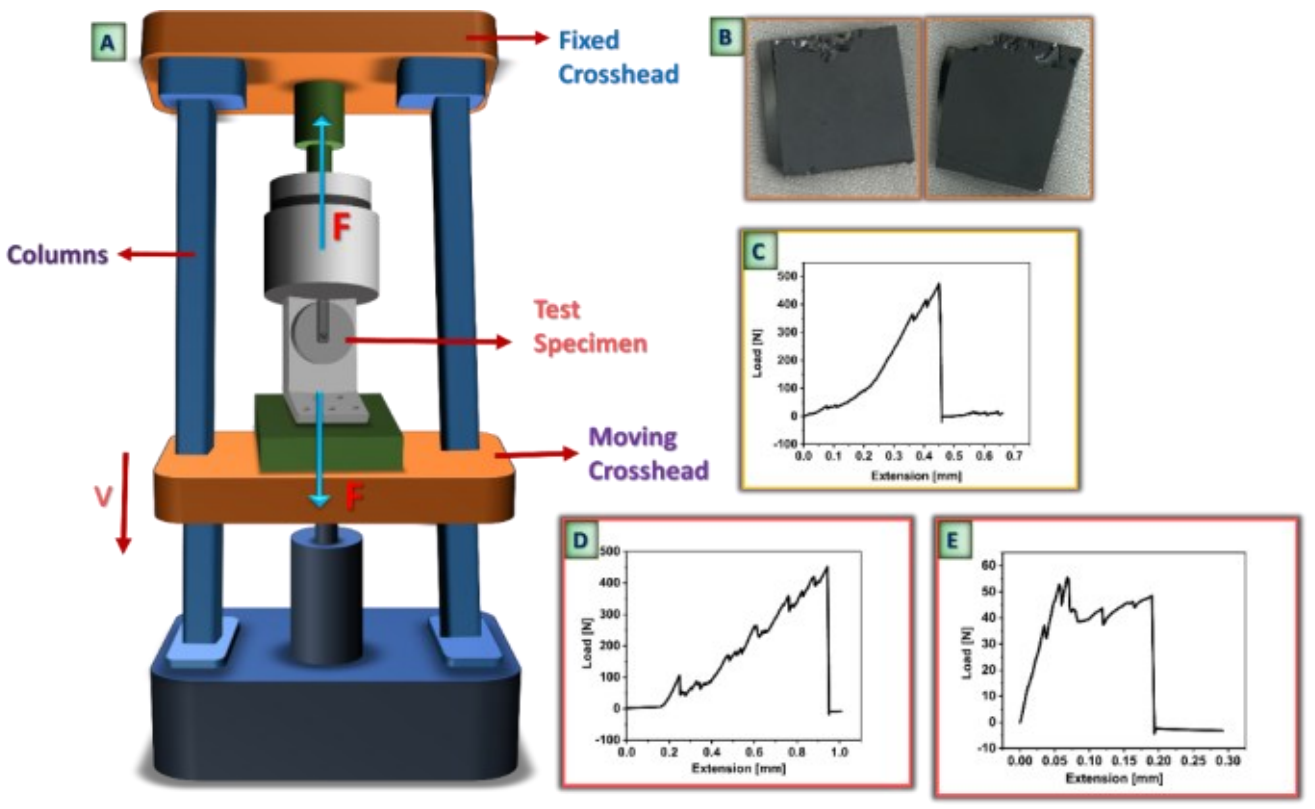

Figure 7. Bond strength measurement (A) Setup for a tensile shear test. (B) Diced bonded samples after the shear test analysis. (C) Tensile shear test of an Au Passivated Ta-Ta bonded sample. (D) Bond strengths were obtained with pause durations of 7 days for the Au passivated Ta-Ta Bonded sample and (E) Bond strength after a 7-day pause without the passivated Ta-Ta Bonded sample.

## V. CONCLUSION

This study presents successful Ta–Ta thermocompression bonding with a 3 nm Au passivation layer at a reduced thermal budget of 300 °C, significantly lower than the previously reported 500 °C. The ultrathin Au passivation layer effectively prevents Ta oxidation and reduces surface roughness to 0.177 nm, as confirmed by XPS, contact angle, and AFM measurements. FESEM and TEM analyses confirm a uniform, defect-free interface. Shear tests show a strong bond strength of 169 MPa, which retains 161 MPa even after 7 days, surpassing that of unpassivated samples, confirming high-quality bonding. This approach demonstrates low-temperature, low-pressure blanket Ta-Ta thermocompression bonding, which widens the way for scalable quantum computing applications at low processing temperatures.

## VI. ACKNOWLEDGEMENT

The author, Mr. Harsh Mishra, acknowledges the financial support provided by the TCS Research Fellowship, as well as the technical assistance from the Nano-X Fabrication Lab and the Characterization Facility at the Indian Institute of Technology Hyderabad.